
%
%
%
\def\unredoffs{} \def\redoffs{\voffset=-.31truein\hoffset=-.48truein}
\def\speclscape{}
%
%
%
%
%
\newbox\leftpage \newdimen\fullhsize \newdimen\hstitle \newdimen\hsbody
\tolerance=1000\hfuzz=2pt
\catcode`\@=11 
\ifx\hyperdef\UNd@FiNeD\def\hyperdef#1#2#3#4{#4}\def\hyperref#1#2#3#4{#4}\fi
\def\bigans{b }
\def\answ{b }
%
\ifx\answ\bigans\message{(This will come out unreduced.}
\magnification=1200\unredoffs\baselineskip=16pt plus 2pt minus 1pt
\hsbody=\hsize \hstitle=\hsize 
\else\message{(This will be reduced.} \let\l@r=L
\magnification=1000\baselineskip=16pt plus 2pt minus 1pt \vsize=7truein
\redoffs \hstitle=8truein\hsbody=4.75truein\fullhsize=10truein\hsize=\hsbody
\output={\ifnum\pageno=0 
  \shipout\vbox{\speclscape{\hsize\fullhsize\makeheadline}
    \hbox to \fullhsize{\hfill\pagebody\hfill}}\advancepageno
  \else
  \almostshipout{\leftline{\vbox{\pagebody\makefootline}}}\advancepageno
  \fi}
\def\almostshipout#1{\if L\l@r \count1=1 \message{[\the\count0.\the\count1]}
      \global\setbox\leftpage=#1 \global\let\l@r=R
 \else \count1=2
  \shipout\vbox{\speclscape{\hsize\fullhsize\makeheadline}
      \hbox to\fullhsize{\box\leftpage\hfil#1}}  \global\let\l@r=L\fi}
\fi
%
\newcount\yearltd\yearltd=\year\advance\yearltd by -1900

\def\Title#1#2{\nopagenumbers\abstractfont\hsize=\hstitle\rightline{#1}%
\vskip 1in\centerline{\titlefont #2}\abstractfont\vskip .5in\pageno=0}
\def\Date#1{\vfill\leftline{#1}\tenpoint\supereject\global\hsize=\hsbody%
\footline={\hss\tenrm\hyperdef\hypernoname{page}\folio\folio\hss}}%
%

\def\draftmode{\message{ DRAFTMODE }\def\draftdate{{\rm preliminary draft:
\number\month/\number\day/\number\yearltd\ \ \hourmin}}%
\headline={\hfil\draftdate}\writelabels\baselineskip=20pt plus 2pt minus 2pt
 {\count255=\time\divide\count255 by 60 \xdef\hourmin{\number\count255}
  \multiply\count255 by-60\advance\count255 by\time
  \xdef\hourmin{\hourmin:\ifnum\count255<10 0\fi\the\count255}}}
\def\nolabels{\def\wrlabeL##1{}\def\eqlabeL##1{}\def\reflabeL##1{}}
\def\writelabels{\def\wrlabeL##1{\leavevmode\vadjust{\rlap{\smash%
{\line{{\escapechar=` \hfill\rlap{\sevenrm\hskip.03in\string##1}}}}}}}%
\def\eqlabeL##1{{\escapechar-1\rlap{\sevenrm\hskip.05in\string##1}}}%
\def\reflabeL##1{\noexpand\llap{\noexpand\sevenrm\string\string\string##1}}}
\nolabels
%
\global\newcount\secno \global\secno=0
\global\newcount\meqno \global\meqno=1
\def\s@csym{}
\def\newsec#1{\global\advance\secno by1%
{\toks0{#1}\message{(\the\secno. \the\toks0)}}%
\global\subsecno=0\eqnres@t\let\s@csym\secsym\xdef\secn@m{\the\secno}\noindent
{\bf\hyperdef\hypernoname{section}{\the\secno}{\the\secno.} #1}%
\writetoca{{\string\hyperref{}{section}{\the\secno}{\the\secno.}} {#1}}%
\par\nobreak\medskip\nobreak}
\def\eqnres@t{\xdef\secsym{\the\secno.}\global\meqno=1\bigbreak\bigskip}
\def\sequentialequations{\def\eqnres@t{\bigbreak}}\xdef\secsym{}
\global\newcount\subsecno \global\subsecno=0
\def\subsec#1{\global\advance\subsecno by1%
{\toks0{#1}\message{(\s@csym\the\subsecno. \the\toks0)}}%
\ifnum\lastpenalty>9000\else\bigbreak\fi
\noindent{\it\hyperdef\hypernoname{subsection}{\secn@m.\the\subsecno}%
{\secn@m.\the\subsecno.} #1}\writetoca{\string\quad
{\string\hyperref{}{subsection}{\secn@m.\the\subsecno}{\secn@m.\the\subsecno.}}
{#1}}\par\nobreak\medskip\nobreak}
\def\appendix#1#2{\global\meqno=1\global\subsecno=0\xdef\secsym{\hbox{#1.}}%
\bigbreak\bigskip\noindent{\bf Appendix \hyperdef\hypernoname{appendix}{#1}%
{#1.} #2}{\toks0{(#1. #2)}\message{\the\toks0}}%
\xdef\s@csym{#1.}\xdef\secn@m{#1}%
\writetoca{\string\hyperref{}{appendix}{#1}{Appendix {#1.}} {#2}}%
\par\nobreak\medskip\nobreak}
%
%
\def\checkm@de#1#2{\ifmmode{\def\f@rst##1{##1}\hyperdef\hypernoname{equation}%
{#1}{#2}}\else\hyperref{}{equation}{#1}{#2}\fi}
\def\eqnn#1{\DefWarn#1\xdef #1{(\noexpand\relax\noexpand\checkm@de%
{\s@csym\the\meqno}{\secsym\the\meqno})}%
\wrlabeL#1\writedef{#1\leftbracket#1}\global\advance\meqno by1}
\def\f@rst#1{\c@t#1a\em@ark}\def\c@t#1#2\em@ark{#1}
\def\eqna#1{\DefWarn#1\wrlabeL{#1$\{\}$}%
\xdef #1##1{(\noexpand\relax\noexpand\checkm@de%
{\s@csym\the\meqno\noexpand\f@rst{##1}}{\hbox{$\secsym\the\meqno##1$}})}
\writedef{#1\numbersign1\leftbracket#1{\numbersign1}}\global\advance\meqno by1}
\def\eqn#1#2{\DefWarn#1%
\xdef #1{(\noexpand\hyperref{}{equation}{\s@csym\the\meqno}%
{\secsym\the\meqno})}$$#2\eqno(\hyperdef\hypernoname{equation}%
{\s@csym\the\meqno}{\secsym\the\meqno})\eqlabeL#1$$%
\writedef{#1\leftbracket#1}\global\advance\meqno by1}
\def\xeqn{\expandafter\xe@n}\def\xe@n(#1){#1}
\def\xeqna#1{\expandafter\xe@n#1}
\def\eqns#1{(\e@ns #1{\hbox{}})}
\def\e@ns#1{\ifx\UNd@FiNeD#1\message{eqnlabel \string#1 is undefined.}%
\xdef#1{(?.?)}\fi{\let\hyperref=\relax\xdef\next{#1}}%
\ifx\next\em@rk\def\next{}\else%
\ifx\next#1\xeqn#1\else\def\n@xt{#1}\ifx\n@xt\next#1\else\xeqna#1\fi
\fi\let\next=\e@ns\fi\next}

\def\DefWarn#1{\ifx\UNd@FiNeD#1\else
\immediate\write16{*** WARNING: the label \string#1 is already defined ***}\fi}
%
\newskip\footskip\footskip14pt plus 1pt minus 1pt 
\def\footnotefont{\ninepoint}\def\f@t#1{\footnotefont #1\@foot}
\def\f@@t{\baselineskip\footskip\bgroup\footnotefont\aftergroup\@foot\let\next}
\setbox\strutbox=\hbox{\vrule height9.5pt depth4.5pt width0pt}
\global\newcount\ftno \global\ftno=0
\def\foot{\global\advance\ftno by1\def\foot@rg{\hyperref{}{footnote}%
{\the\ftno}{\the\ftno}\xdef\foot@rg{\noexpand\hyperdef\noexpand\hypernoname%
{footnote}{\the\ftno}{\the\ftno}}}\footnote{$^{\foot@rg}$}}
%
\newwrite\ftfile
\def\footend{\def\foot{\global\advance\ftno by1\chardef\wfile=\ftfile
\hyperref{}{footnote}{\the\ftno}{$^{\the\ftno}$}%
\ifnum\ftno=1\immediate\openout\ftfile=\jobname.fts\fi%
\immediate\write\ftfile{\noexpand\smallskip%
\noexpand\item{\noexpand\hyperdef\noexpand\hypernoname{footnote}
{\the\ftno}{f\the\ftno}:\ }\pctsign}\findarg}%
\def\footatend{\vfill\eject\immediate\closeout\ftfile{\parindent=20pt
\centerline{\bf Footnotes}\nobreak\bigskip\input \jobname.fts }}}
\def\footatend{}
%
%
\global\newcount\refno \global\refno=1
\newwrite\rfile
\def\ref{[\hyperref{}{reference}{\the\refno}{\the\refno}]\nref}
\def\nref#1{\DefWarn#1%
\xdef#1{[\noexpand\hyperref{}{reference}{\the\refno}{\the\refno}]}%
\writedef{#1\leftbracket#1}%
\ifnum\refno=1\immediate\openout\rfile=\jobname.refs\fi
\chardef\wfile=\rfile\immediate\write\rfile{\noexpand\item{[\noexpand\hyperdef%
\noexpand\hypernoname{reference}{\the\refno}{\the\refno}]\ }%
\reflabeL{#1\hskip.31in}\pctsign}\global\advance\refno by1\findarg}
\def\findarg#1#{\begingroup\obeylines\newlinechar=`\^^M\pass@rg}
{\obeylines\gdef\pass@rg#1{\writ@line\relax #1^^M\hbox{}^^M}%
\gdef\writ@line#1^^M{\expandafter\toks0\expandafter{\striprel@x #1}%
\edef\next{\the\toks0}\ifx\next\em@rk\let\next=\endgroup\else\ifx\next\empty%
\else\immediate\write\wfile{\the\toks0}\fi\let\next=\writ@line\fi\next\relax}}
\def\striprel@x#1{} \def\em@rk{\hbox{}}
\def\lref{\begingroup\obeylines\lr@f}
\def\lr@f#1#2{\DefWarn#1\gdef#1{\let#1=\UNd@FiNeD\ref#1{#2}}\endgroup\unskip}

\def\addref#1{\immediate\write\rfile{\noexpand\item{}#1}} 
\def\listrefs{\footatend\vfill\supereject\immediate\closeout\rfile\writestoppt
\baselineskip=\footskip\centerline{{\bf References}}\bigskip{\parindent=20pt%
\frenchspacing\escapechar=` \input \jobname.refs\vfill\eject}\nonfrenchspacing}
\def\startrefs#1{\immediate\openout\rfile=\jobname.refs\refno=#1}
\def\xref{\expandafter\xr@f}\def\xr@f[#1]{#1}
\def\refs#1{\count255=1[\r@fs #1{\hbox{}}]}
\def\r@fs#1{\ifx\UNd@FiNeD#1\message{reflabel \string#1 is undefined.}%
\nref#1{need to supply reference \string#1.}\fi%
\vphantom{\hphantom{#1}}{\let\hyperref=\relax\xdef\next{#1}}%
\ifx\next\em@rk\def\next{}%
\else\ifx\next#1\ifodd\count255\relax\xref#1\count255=0\fi%
\else#1\count255=1\fi\let\next=\r@fs\fi\next}
%

%
\newwrite\ffile\global\newcount\figno \global\figno=1
\def\fig{fig.~\hyperref{}{figure}{\the\figno}{\the\figno}\nfig}
\def\nfig#1{\DefWarn#1%
\xdef#1{fig.~\noexpand\hyperref{}{figure}{\the\figno}{\the\figno}}%
\writedef{#1\leftbracket fig.\noexpand~\xfig#1}%
\ifnum\figno=1\immediate\openout\ffile=\jobname.figs\fi\chardef\wfile=\ffile%
{\let\hyperref=\relax
\immediate\write\ffile{\noexpand\medskip\noexpand\item{Fig.\ %
\noexpand\hyperdef\noexpand\hypernoname{figure}{\the\figno}{\the\figno}. }
\reflabeL{#1\hskip.55in}\pctsign}}\global\advance\figno by1\findarg}
\def\listfigs{\vfill\eject\immediate\closeout\ffile{\parindent40pt
\baselineskip14pt\centerline{{\bf Figure Captions}}\nobreak\medskip
\escapechar=` \input \jobname.figs\vfill\eject}}
\def\xfig{\expandafter\xf@g}\def\xf@g fig.\penalty\@M\ {}
\def\figs#1{figs.~\f@gs #1{\hbox{}}}
\def\f@gs#1{{\let\hyperref=\relax\xdef\next{#1}}\ifx\next\em@rk\def\next{}\else
\ifx\next#1\xfig #1\else#1\fi\let\next=\f@gs\fi\next}
\def\figin{\epsfcheck\figin}\def\figins{\epsfcheck\figins}
\def\epsfcheck{\ifx\epsfbox\UNd@FiNeD
\message{(NO epsf.tex, FIGURES WILL BE IGNORED)}
\gdef\figin##1{\vskip2in}\gdef\figins##1{\hskip.5in}
\else\message{(FIGURES WILL BE INCLUDED)}%
\gdef\figin##1{##1}\gdef\figins##1{##1}\fi}
\def\DefWarn#1{}
\def\figinsert{\goodbreak\midinsert}
\def\ifig#1#2#3{\DefWarn#1\xdef#1{fig.~\noexpand\hyperref{}{figure}%
{\the\figno}{\the\figno}}\writedef{#1\leftbracket fig.\noexpand~\xfig#1}%
\figinsert\figin{\centerline{#3}}\medskip\centerline{\vbox{\baselineskip12pt
\advance\hsize by -1truein\noindent\wrlabeL{#1=#1}\footnotefont%
{\bf Fig.~\hyperdef\hypernoname{figure}{\the\figno}{\the\figno}:} #2}}
\bigskip\endinsert\global\advance\figno by1}
\newwrite\lfile
{\escapechar-1\xdef\pctsign{\string\%}\xdef\leftbracket{\string\{}
\xdef\rightbracket{\string\}}\xdef\numbersign{\string\#}}
\def\writedefs{\immediate\openout\lfile=\jobname.defs \def\writedef##1{%
{\let\hyperref=\relax\let\hyperdef=\relax\let\hypernoname=\relax
 \immediate\write\lfile{\string\def\string##1\rightbracket}}}}%
\def\writestop{\def\writestoppt{\immediate\write\lfile{\string\pageno
 \the\pageno\string\startrefs\leftbracket\the\refno\rightbracket
 \string\def\string\secsym\leftbracket\secsym\rightbracket
 \string\secno\the\secno\string\meqno\the\meqno}\immediate\closeout\lfile}}
\def\writestoppt{}\def\writedef#1{}
\def\seclab#1{\DefWarn#1%
\xdef #1{\noexpand\hyperref{}{section}{\the\secno}{\the\secno}}%
\writedef{#1\leftbracket#1}\wrlabeL{#1=#1}}
\def\subseclab#1{\DefWarn#1%
\xdef #1{\noexpand\hyperref{}{subsection}{\secn@m.\the\subsecno}%
{\secn@m.\the\subsecno}}\writedef{#1\leftbracket#1}\wrlabeL{#1=#1}}
\def\applab#1{\DefWarn#1%
\xdef #1{\noexpand\hyperref{}{appendix}{\secn@m}{\secn@m}}%
\writedef{#1\leftbracket#1}\wrlabeL{#1=#1}}
\newwrite\tfile \def\writetoca#1{}
\def\leaderfill{\leaders\hbox to 1em{\hss.\hss}\hfill}
\def\writetoc{\immediate\openout\tfile=\jobname.toc
   \def\writetoca##1{{\edef\next{\write\tfile{\noindent ##1
   \string\leaderfill {\string\hyperref{}{page}{\noexpand\number\pageno}%
                       {\noexpand\number\pageno}} \par}}\next}}}
\newread\ch@ckfile
\def\listtoc{\immediate\closeout\tfile\immediate\openin\ch@ckfile=\jobname.toc
\ifeof\ch@ckfile\message{no file \jobname.toc, no table of contents this pass}%
\else\closein\ch@ckfile\centerline{\bf Contents}\nobreak\medskip%
{\baselineskip=12pt\footnotefont\parskip=0pt\catcode`\@=11\input\jobname.toc
\catcode`\@=12\bigbreak\bigskip}\fi}
\catcode`\@=12 
%
\edef\tfontsize{\ifx\answ\bigans scaled\magstep3\else scaled\magstep4\fi}
\font\titlerm=cmr10 \tfontsize \font\titlerms=cmr7 \tfontsize
\font\titlermss=cmr5 \tfontsize \font\titlei=cmmi10 \tfontsize
\font\titleis=cmmi7 \tfontsize \font\titleiss=cmmi5 \tfontsize
\font\titlesy=cmsy10 \tfontsize \font\titlesys=cmsy7 \tfontsize
\font\titlesyss=cmsy5 \tfontsize \font\titleit=cmti10 \tfontsize
\skewchar\titlei='177 \skewchar\titleis='177 \skewchar\titleiss='177
\skewchar\titlesy='60 \skewchar\titlesys='60 \skewchar\titlesyss='60
\def\titlefont{\def\rm{\fam0\titlerm}
\textfont0=\titlerm \scriptfont0=\titlerms \scriptscriptfont0=\titlermss
\textfont1=\titlei \scriptfont1=\titleis \scriptscriptfont1=\titleiss
\textfont2=\titlesy \scriptfont2=\titlesys \scriptscriptfont2=\titlesyss
\textfont\itfam=\titleit \def\it{\fam\itfam\titleit}\rm}
 \ifx\answ\bigans\else scaled\magstep1\fi
\ifx\answ\bigans\def\abstractfont{\tenpoint}\else
\font\absit=cmti10 scaled \magstep1
\font\abssl=cmsl10 scaled \magstep1
\font\absrm=cmr10 scaled\magstep1 \font\absrms=cmr7 scaled\magstep1
\font\absrmss=cmr5 scaled\magstep1 \font\absi=cmmi10 scaled\magstep1
\font\absis=cmmi7 scaled\magstep1 \font\absiss=cmmi5 scaled\magstep1
\font\abssy=cmsy10 scaled\magstep1 \font\abssys=cmsy7 scaled\magstep1
\font\abssyss=cmsy5 scaled\magstep1 \font\absbf=cmbx10 scaled\magstep1
\skewchar\absi='177 \skewchar\absis='177 \skewchar\absiss='177
\skewchar\abssy='60 \skewchar\abssys='60 \skewchar\abssyss='60
\def\abstractfont{\def\rm{\fam0\absrm}
\textfont0=\absrm \scriptfont0=\absrms \scriptscriptfont0=\absrmss
\textfont1=\absi \scriptfont1=\absis \scriptscriptfont1=\absiss
\textfont2=\abssy \scriptfont2=\abssys \scriptscriptfont2=\abssyss
\textfont\itfam=\absit \def\it{\fam\itfam\absit}\def\footnotefont{\tenpoint}%
\textfont\slfam=\abssl \def\sl{\fam\slfam\abssl}%
\textfont\bffam=\absbf \def\bf{\fam\bffam\absbf}\rm}\fi
\def\tenpoint{\def\rm{\fam0\tenrm}
\textfont0=\tenrm \scriptfont0=\sevenrm \scriptscriptfont0=\fiverm
\textfont1=\teni  \scriptfont1=\seveni  \scriptscriptfont1=\fivei
\textfont2=\tensy \scriptfont2=\sevensy \scriptscriptfont2=\fivesy
\textfont\itfam=\tenit \def\it{\fam\itfam\tenit}\def\footnotefont{\ninepoint}%
\textfont\bffam=\tenbf \def\bf{\fam\bffam\tenbf}\def\sl{\fam\slfam\tensl}\rm}
\font\ninerm=cmr9 \font\sixrm=cmr6 \font\ninei=cmmi9 \font\sixi=cmmi6
\font\ninesy=cmsy9 \font\sixsy=cmsy6 \font\ninebf=cmbx9
\font\nineit=cmti9 \font\ninesl=cmsl9 \skewchar\ninei='177
\skewchar\sixi='177 \skewchar\ninesy='60 \skewchar\sixsy='60
\def\ninepoint{\def\rm{\fam0\ninerm}
\textfont0=\ninerm \scriptfont0=\sixrm \scriptscriptfont0=\fiverm
\textfont1=\ninei \scriptfont1=\sixi \scriptscriptfont1=\fivei
\textfont2=\ninesy \scriptfont2=\sixsy \scriptscriptfont2=\fivesy
\textfont\itfam=\ninei \def\it{\fam\itfam\nineit}\def\sl{\fam\slfam\ninesl}%
\textfont\bffam=\ninebf \def\bf{\fam\bffam\ninebf}\rm}
%
%

\hyphenation{anom-aly anom-alies coun-ter-term coun-ter-terms}
\def\inv{^{\raise.15ex\hbox{${\scriptscriptstyle -}$}\kern-.05em 1}}

\def\Dsl{\,\raise.15ex\hbox{/}\mkern-13.5mu D} 
\def\dsl{\raise.15ex\hbox{/}\kern-.57em\partial}

\def\tr{{\rm tr}} 
\def\lspace{\ifx\answ\bigans{}\else\qquad\fi}
\def\lbspace{\ifx\answ\bigans{}\else\hskip-.2in\fi} 
\def\boxeqn#1{\vcenter{\vbox{\hrule\hbox{\vrule\kern3pt\vbox{\kern3pt
	\hbox{${\displaystyle #1}$}\kern3pt}\kern3pt\vrule}\hrule}}}
\def\mbox#1#2{\vcenter{\hrule \hbox{\vrule height#2in
		\kern#1in \vrule} \hrule}}  
%

\def\darr#1{\raise1.5ex\hbox{$\leftrightarrow$}\mkern-16.5mu #1}

\def\half{{\textstyle{1\over2}}} 
\def\roughly#1{\raise.3ex\hbox{$#1$\kern-.75em\lower1ex\hbox{$\sim$}}}

\input epsf
\newif\ifdraft\draftfalse
\newif\ifinter\interfalse
\ifdraft\draftmode\else\interfalse\fi
\def\journal#1&#2(#3){\unskip, \sl #1\ \bf #2 \rm(19#3) }
\def\andjournal#1&#2(#3){\sl #1~\bf #2 \rm (19#3) }

\def\frac#1#2{{#1\over#2}}

\def\half{\frac12}

\def\inbar{\,\vrule height1.5ex width.4pt depth0pt}
\def\IC{\relax\hbox{$\inbar\kern-.3em{\rm C}$}}
\def\IR{\relax{\rm I\kern-.18em R}}
\def\IP{\relax{\rm I\kern-.18em P}}

%
%


%
\catcode`\@=11
\def\slash#1{\mathord{\mathpalette\c@ncel{#1}}}
\overfullrule=0pt

\def\MM{{\cal M}}
\def\NN{{\cal N}}

\def\underrel#1\over#2{\mathrel{\mathop{\kern\z@#1}\limits_{#2}}}

\catcode`\@=12


%

\def\tr{{\rm tr}}


\def\[{[}
\def\]{]}

\def\comment#1{ }

%
\def\draftnote#1{\ifdraft{\baselineskip2ex
                 \vbox{\kern1em\hrule\hbox{\vrule\kern1em\vbox{\kern1ex
                 \noindent \underbar{NOTE}: #1
             \vskip1ex}\kern1em\vrule}\hrule}}\fi}
\def\internote#1{\ifinter{\baselineskip2ex
                 \vbox{\kern1em\hrule\hbox{\vrule\kern1em\vbox{\kern1ex
                 \noindent \underbar{Internal Note}: #1
             \vskip1ex}\kern1em\vrule}\hrule}}\fi}

%
%



%
%
%
%

%

\def\inv{^{-1}}


\def\al{\tilde a}
\def\cl{\tilde c}

\def\x{{\tilde x}}
\def\tr{{\rm tr}}

\def\ra{{\rightarrow}}

\def\y{{\tilde y}}

\lref\KutasovIY{
  D.~Kutasov, A.~Parnachev and D.~A.~Sahakyan,
  ``Central charges and U(1)R symmetries in N = 1 super Yang-Mills,''
  JHEP {\bf 0311}, 013 (2003)
  [arXiv:hep-th/0308071].
}

\lref\IntriligatorMI{
  K.~A.~Intriligator and B.~Wecht,
  ``RG fixed points and flows in SQCD with adjoints,''
  Nucl.\ Phys.\  B {\bf 677}, 223 (2004)
  [arXiv:hep-th/0309201].
}

\lref\Brodie{
  J.~H.~Brodie,
  ``Duality in supersymmetric SU(N/c) gauge theory with two adjoint chiral
  superfields,''
  Nucl.\ Phys.\  B {\bf 478}, 123 (1996)
  [arXiv:hep-th/9605232].
}

\lref\LatorreEA{
  J.~I.~Latorre and H.~Osborn,
  ``Modified weak energy condition for the energy momentum tensor in  quantum
  field theory,''
  Nucl.\ Phys.\  B {\bf 511}, 737 (1998)
  [arXiv:hep-th/9703196].
}

\lref\IntriligatorJJ{
K.~Intriligator and B.~Wecht,
``The exact superconformal R-symmetry maximizes a,''
arXiv:hep-th/0304128.
}

\lref\ShapereZF{
  A.~D.~Shapere and Y.~Tachikawa,
  ``Central charges of N=2 superconformal field theories in four dimensions,''
  JHEP {\bf 0809}, 109 (2008)
  [arXiv:0804.1957 [hep-th]].
}

\lref\CardyCW{
J.~L.~Cardy,
``Is There A C Theorem In Four-Dimensions?,''
Phys.\ Lett.\ B {\bf 215}, 749 (1988).
}
\lref\JackEB{
I.~Jack and H.~Osborn,
``Analogs For The C Theorem For Four-Dimensional Renormalizable Field Theories,''
Nucl.\ Phys.\ B {\bf 343}, 647 (1990).
}
\lref\CappelliYC{
A.~Cappelli, D.~Friedan and J.~I.~Latorre,
``C Theorem And Spectral Representation,''
Nucl.\ Phys.\ B {\bf 352}, 616 (1991).
}
\lref\CappelliKE{
A.~Cappelli, J.~I.~Latorre and X.~Vilasis-Cardona,
``Renormalization group patterns and C theorem in more than two-dimensions,''
Nucl.\ Phys.\ B {\bf 376}, 510 (1992)
[arXiv:hep-th/9109041].
}
\lref\BastianelliVV{
F.~Bastianelli,
``Tests for C-theorems in 4D,''
Phys.\ Lett.\ B {\bf 369}, 249 (1996)
[arXiv:hep-th/9511065].
}

\lref\ForteDX{
S.~Forte and J.~I.~Latorre,
``A proof of the irreversibility of renormalization group flows in four  dimensions,''
Nucl.\ Phys.\ B {\bf 535}, 709 (1998)
[arXiv:hep-th/9805015].
}

\lref\AnselmiYS{
D.~Anselmi, J.~Erlich, D.~Z.~Freedman and A.~A.~Johansen,
``Positivity constraints on anomalies in supersymmetric gauge theories,''
Phys.\ Rev.\ D {\bf 57}, 7570 (1998)
[arXiv:hep-th/9711035].
}

\lref\AnselmiAM{
D.~Anselmi, D.~Z.~Freedman, M.~T.~Grisaru and A.~A.~Johansen,
``Nonperturbative formulas for central functions of supersymmetric gauge  theories,''
Nucl.\ Phys.\ B {\bf 526}, 543 (1998)
[arXiv:hep-th/9708042].
}

\lref\AnselmiUK{
D.~Anselmi,
``Quantum irreversibility in arbitrary dimension,''
Nucl.\ Phys.\ B {\bf 567}, 331 (2000)
[arXiv:hep-th/9905005].
}
\lref\CappelliDV{
A.~Cappelli, G.~D'Appollonio, R.~Guida and N.~Magnoli,
``On the c-theorem in more than two dimensions,''
arXiv:hep-th/0009119.
}

\lref\SeibergPQ{
N.~Seiberg,
``Electric - magnetic duality in supersymmetric nonAbelian gauge theories,''
Nucl.\ Phys.\ B {\bf 435}, 129 (1995)
[arXiv:hep-th/9411149].
}

\lref\KutasovVE{
D.~Kutasov,
``A Comment on duality in N=1 supersymmetric nonAbelian gauge theories,''
Phys.\ Lett.\ B {\bf 351}, 230 (1995)
[arXiv:hep-th/9503086].
}
\lref\KutasovNP{
D.~Kutasov and A.~Schwimmer,
``On duality in supersymmetric Yang-Mills theory,''
Phys.\ Lett.\ B {\bf 354}, 315 (1995)
[arXiv:hep-th/9505004].
}
\lref\KutasovSS{
D.~Kutasov, A.~Schwimmer and N.~Seiberg,
Nucl.\ Phys.\ B {\bf 459}, 455 (1996)
[arXiv:hep-th/9510222].
}
\lref\AffleckMK{
I.~Affleck, M.~Dine and N.~Seiberg,
``Dynamical Supersymmetry Breaking In Supersymmetric QCD,''
Nucl.\ Phys.\ B {\bf 241}, 493 (1984).
}
\lref\AffleckXZ{
I.~Affleck, M.~Dine and N.~Seiberg,
``Dynamical Supersymmetry Breaking In Four-Dimensions And Its Phenomenological Implications,''
Nucl.\ Phys.\ B {\bf 256}, 557 (1985).
}
\lref\ZamolodchikovGT{
A.~B.~Zamolodchikov,
JETP Lett.\  {\bf 43}, 730 (1986)
[Pisma Zh.\ Eksp.\ Teor.\ Fiz.\  {\bf 43}, 565 (1986)].
}

\lref\HofmanAR{
  D.~M.~Hofman and J.~Maldacena,
  ``Conformal collider physics: Energy and charge correlations,''
  JHEP {\bf 0805}, 012 (2008)
  [arXiv:0803.1467 [hep-th]].
}

\lref\AnselmiAM{
  D.~Anselmi, D.~Z.~Freedman, M.~T.~Grisaru and A.~A.~Johansen,
  Nucl.\ Phys.\  B {\bf 526}, 543 (1998)
  [arXiv:hep-th/9708042].
}

\lref\ShapereUN{
  A.~D.~Shapere and Y.~Tachikawa,
  ``A counterexample to the 'a-theorem',''
  arXiv:0809.3238 [hep-th].
}

\lref\SeibergPQ{
  N.~Seiberg,
  ``Electric - magnetic duality in supersymmetric nonAbelian gauge theories,''
  Nucl.\ Phys.\  B {\bf 435}, 129 (1995)
  [arXiv:hep-th/9411149].
}

\lref\ZamolodchikovGT{
  A.~B.~Zamolodchikov,
  ``Irreversibility of the Flux of the Renormalization Group in a 2D Field
  Theory,''
  JETP Lett.\  {\bf 43}, 730 (1986)
  [Pisma Zh.\ Eksp.\ Teor.\ Fiz.\  {\bf 43}, 565 (1986)].
}

\lref\KutasovUX{
  D.~Kutasov,
  ``New results on the 'a-theorem' in four dimensional supersymmetric field
  theory,''
  arXiv:hep-th/0312098.
}

\lref\KutasovXU{
  D.~Kutasov and A.~Schwimmer,
  ``Lagrange multipliers and couplings in supersymmetric field theory,''
  Nucl.\ Phys.\  B {\bf 702}, 369 (2004)
  [arXiv:hep-th/0409029].
}

\lref\CsakiUJ{
  C.~Csaki, P.~Meade and J.~Terning,
  ``A mixed phase of SUSY gauge theories from a-maximization,''
  JHEP {\bf 0404}, 040 (2004)
  [arXiv:hep-th/0403062].
}

\lref\BarnesJJ{
  E.~Barnes, K.~A.~Intriligator, B.~Wecht and J.~Wright,
  ``Evidence for the strongest version of the 4d a-theorem, via  a-maximization
  along RG flows,''
  Nucl.\ Phys.\  B {\bf 702}, 131 (2004)
  [arXiv:hep-th/0408156].
}

\lref\KutasovNP{
  D.~Kutasov and A.~Schwimmer,
  ``On duality in supersymmetric Yang-Mills theory,''
  Phys.\ Lett.\  B {\bf 354}, 315 (1995)
  [arXiv:hep-th/9505004].
}

\Title{\vbox{\baselineskip12pt
\hbox{YITP-SB-08-46}
}}
{\vbox{\centerline{Comments on Bounds on Central Charges in $\NN=1$ }
\vskip.06in
\centerline{Superconformal Theories}
\vskip.06in
}}
\centerline{Andrei Parnachev and Shlomo S. Razamat}
\bigskip
\centerline{{\it C.N.Yang Institute for Theoretical Physics }}
\centerline{\it Stony Brook University }
\centerline{\it Stony Brook, NY 11794-3840, USA}
\vskip.1in \vskip.1in \centerline{\bf Abstract}
\noindent
The ratio of central charges in four-dimensional $CFT$s has been suggested
by Hofman and Maldacena to lie within an interval whose boundaries are fixed by the number of
supersymmetries.
We compute this ratio for a set of interacting $\NN=1$ superconformal field theories 
which arise as $RG$  fixed points of supersymmetric Yang-Mills theories with adjoint and fundamental
matter.
We do not find violations of the proposed bounds, which appear to be  saturated
by free field theories.

\vfill

\Date{December 2008}


\newsec{Introduction and Summary.}
\noindent 
Weyl anomaly in four-dimensional conformal field theories contains 
a term proportional to the Euler density, and a square of the Weyl tensor.
The properly normalized coefficients in front of these terms are the
$a$ and $c$ central charges of the conformal group.
Their values can be determined from the three and two-point functions of stress-energy
tensor.
In the $\NN=1$ superconformal theories one can relate $a$ and $c$ to the 't Hooft anomalies
involving the $R$-current \refs{\AnselmiAM,\AnselmiYS}.
Hence, the values of $a$ and $c$ can be computed through
\eqn\acRrel{a=\frac{3}{32}\left(3\,Tr\,R^3-Tr\,R\right),\qquad c=\frac{1}{32}\left(9\,Tr\,R^3-5\,Tr\,R\right).}
provided one can identify the correct 
$U(1)_R$ symmetry in the $IR$.
The problem is that any global $U(1)$ which commutes with the superconformal group
can become part of the $U(1)_R$ symmetry in the $IR$.
The solution, worked out by Intriligator and Wecht \IntriligatorJJ, is that the 
correct  $U(1)_R$ symmetry maximizes  $a$.
If the $IR$ fixed point is not too strongly coupled
 the maximization of $a$  allows computing
 the values of the central charges in a variety of interesting
examples.
Otherwise, the appearance of ``accidental symmetries'' which are not visible
in the UV would complicate the situation.

The central charges $a$ and $c$ count the degrees of freedom of the $CFT$.
It is therefore natural to ask whether they satisfy any nontrivial constraints.
While positivity of $c$ follows from the unitarity of the theory \refs{\AnselmiAM,\AnselmiYS}, 
other properties have more conjectural nature.
For example, $a$ has been proposed to satisfy an 
analog of Zamolodchikov $c$-theorem \ZamolodchikovGT\ in four dimensions \CardyCW. (See
\refs{\AnselmiYS,\KutasovUX\CsakiUJ\BarnesJJ-\KutasovXU} for some work in this direction
and  \ShapereUN\ which argued that a counterexample exists.)

Recently new bounds on the ratio of central charges in $\NN=1$ superconformal 
theories have been suggested in \HofmanAR.\foot{See \LatorreEA\ for related work.} 
By requiring the positivity of  the  energy flux, \HofmanAR\ derived the following 
inequality for the $\NN=1$ superconformal theories,
\eqn\alphabound{   -{1\over2}\leq\alpha\leq {1\over2},  }
where we define 
\eqn\defalpha{  \alpha={a\over c}-1.  }
There are similar constraints for the $CFT$s with $\NN=0,2$ supersymmetries,
with the $\NN=2$ bound  subsequently proven in \ShapereZF.
Assuming positivity of the energy flux seems natural and can even be proven for
free field theories.
However, there is no proof for a generic  interacting $CFT$.

In this note we compute the values of $\alpha$, as defined
in \defalpha , for a set of interacting $CFT$s studied in \IntriligatorMI.
This includes the IR fixed points of supersymmetric $QCD$ with one and two
adjoints with all possible relevant superpotential deformations.
We use the prescription of  \IntriligatorJJ\ (slightly modified to take
care of decoupled composite fields \KutasovIY) to compute the
central charges.
We do not find any violations of the bound \alphabound; in fact
in all of the interacting examples we consider 
the value of $\alpha$ is negative and larger than $-\half$.

We mostly follow notations and results from  \KutasovIY\ and \IntriligatorMI.
In particular, we will work in the large $N$ limit
\eqn\limit{
N_c>>1;\quad N_f>>1;\quad x\equiv {N_c\over N_f}\,\, =\,\, {\rm fixed}~,
}
and study the quantity $\alpha=a/c-1$  as a function of the continuous parameter $x$. 

The simplest non-trivial example is provided by 
supersymmetric $QCD$, whose 
field  content
consists of a vector superfield in the adjoint of $SU(N_c)$,
and $N_f$ chiral superfields $Q$ and $\tilde Q$ in the (anti)fundamental 
of $SU(N_c)$.
The anomaly cancellation fixes the $R$-charges of the  $Q$ and $\tilde Q$
to be
\eqn\rqsqcd{  R(Q)=R({\tilde Q})=1-x . }
The electric theory is asymptotically free for $x>1/3$, while
the magnetic theory ( with the gauge group $SU(N_f-N_c)$) is asymptotically free for $x<2/3$.
The two theories are Seiberg dual \SeibergPQ\ (flow to the same IR
fixed point).
Using the relations \acRrel\ we compute the value of $\alpha$:
\eqn\alphasqcd{ \alpha={1\over 9 x^2-7}.  }
which interpolates between $-1/3$ and $-1/6$ within the 
conformal window $1/3<x<2/3$, and thus satisfies the bounds \alphabound.
Note that the bound would naively be violated for e.g. $x>\sqrt{5}/3$.
However in this regime electric theory is strongly coupled and we should
trust the magnetic description, which gives free theory in the IR.

One can discuss more general theories by adding adjoint chiral  
superfields. Asymptotic freedom implies that the number of chiral
adjoints will be at most two.
The rest of the paper is organized as follows.
In sections 2 and 3 we consider supersymmetric $QCD$ with one and two adjoints and possible 
relevant superpotentials, and briefly conclude in section 4.


\newsec{$sQCD$ with one adjoint}
\subsec{Vanishing superpotential}
\noindent
The field content of supersymmetric $QCD$ with one adjoint field
consists of a vector and a chiral superfields both transforming in the adjoint of $SU(N_c)$,
and $N_f$ chiral superfields $Q$ and $\tilde Q$ in the (anti)fundamental 
of $SU(N_c)$.
The theory
is asymptotically free for $x>1/2$, and we will restrict our discussion to this regime.

The anomaly cancellation condition implies that the $R$-charges of $X,Q,\tilde Q$
depend on a single parameter.
More precisely, denoting the $R$-charge of $Q$ by $y$, gives 
the $R$-charge of $X$ to be $R(X)=(1-y)/x$.
As found in \IntriligatorJJ, to determine the values of the $R$-charges 
we need to maximize the $a$-function with respect to $y$.
In what follows we will discuss the scaled functions,
\eqn\tildeac{\tilde a=\frac{32}{3}a,\qquad \tilde c=\frac{32}{3}c.}
In the first approximation, the trial $\tilde a$-function is 
\eqn\naiv{
\tilde a^{(0)}(x,y)/N_f^2=6x(y-1)^3-2x(y-1)+3x^2\left({1-y\over x}-1\right)^3-
x^2\left({1-y\over x}-1\right)+2x^2~.}
while the c-function is given by
\eqn\naivc{
\tilde c^{(0)}(x,y)/N_f^2=6x(y-1)^3-{10\over3}x(y-1)+3x^2\left({1-y\over x}-1\right)^3-
{5\over3} x^2\left({1-y\over x}-1\right)+{4\over3}x^2~.}
and 
\eqn\naivac{  \al^{(0)}-\cl^{(0)} = {2\over3} x (y-1).  }
Maximizing \naiv\ with respect to $y$ gives 
\eqn\naivR{
R^{(0)}(Q)=y^{(0)}={3+x(-3-6x+\sqrt{20x^2-1})\over 3-6x^2.}
}
Plugging \naivR\ into \naiv\ and \naivc, one finds the following expression for 
$\alpha$:
\eqn\aattii{\alpha(x)=   {4 x^2-2\over 13-44 x^2 +3\sqrt{20x^2-1}}.   }
which is a monotonic function interpolating between $\alpha(1/2)=-1/8$ and
$\alpha(x\ra\infty)=-1/11$.

As explained in \KutasovIY, this is not the full story as unitarity 
corrections due to decoupling mesons $\MM_j= {\tilde Q} X^{j-1} Q$ need to be taken into account.
In particular, the naive $R$-charges of these superfields,
\eqn\mes{
R({\cal M}_j)=2y+(j-1){1-y\over x},
}
can get lower then the unitarity bound, $R({\cal M}_j)={2\over3}$.
In this situation the infrared $CFT$
splits into an (in general) interacting theory and a decoupled free superfield $\MM_j$.
The correct central charges are obtained by subtracting the contribution 
of such mesons with  the naive $R$-charges \mes\ and adding back the
contribution due to the free fields, see \KutasovIY\ for details.

In practice, it is convenient to introduce 
central charges computed with the assumption that the first $p(=0,1,2,\cdots)$ 
meson fields
$\MM_1,\cdots,\MM_p$ are free:
\eqn\acorrect{  \al^{(p)}=\al^{(0)}+\sum_{j=1}^p \left[{2\over9} -3 R(M_j)+R(M_j)\right],   }
and
\eqn\ccorrect{ \cl^{(p)}=\cl^{(0)}+\sum_{j=1}^p \left[{4\over9} -3 R(M_j)+{5\over3} R(M_j)\right],   }
where $\al^{(0)}$ and $\cl^{(0)}$ are given by \naiv\ and \naivc\ respectively.
Substituting the values of $R(M_j)$ from \mes\ we get e.g.
\eqn\correct{
\eqalign{
&\al^{(p)}(x,y)/N_f^2
=6x(y-1)^3-2x(y-1)+3x^2\left({1-y\over x}-1\right)^3-
x^2\left({1-y\over x}-1\right)+2x^2+\cr
&\cr
&{1\over 9}\sum_{j=1}^p
\left[2- 3\left(2y + (j - 1){1 - y\over x}\right)\right]^2
\left[5 - 3\left(2y + (j - 1){1 - y\over x}\right)\right]~.
}
}
The central charges can then be determined via the following process:
Start with $\al^{(0)}$ and find the
value of $x$ for which the $R$-charge of $\MM_1=\tilde Q Q$ approaches $2/3$. At that
point  ${\cal M}_1$ becomes free and we have to switch to the $\al^{(1)}$ description. 
Then look for the value of $x$ at which ${\cal M}_2$ becomes free
and decouples, switch to $\al^{(2)}$, etc. This process can be obviously 
continued to arbitrarily large $x$.
\midinsert\bigskip{\vbox{{\epsfxsize=3in
        \nobreak
    \centerline{\epsfbox{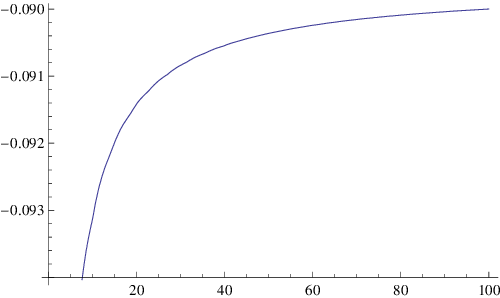}}
        \nobreak\bigskip
    {\raggedright\it \vbox{
{\bf Fig 1.}
{\it  $\alpha$ for $sQCD$ with $W=0$.
}}}}}}
\bigskip\endinsert
\noindent

As shown in \KutasovIY, this procedure is justified because
once the meson decouples its $R$-charge \mes\ stays below the unitarity bound.
We implemented this algorithm using Mathematica and computed the values
of ${\tilde a}(x)$, ${\tilde c}(x)$.
The result for  $\alpha(x)$ is shown in Fig. 1.

\subsec{Adjoint $sQCD$ with polynomial superpotential}
\noindent
Adding the relevant superpotential of the type,
\eqn\wel{
W_k(X)=g_k\tr X^{k+1},
}
to adjoint $sQCD$ induces the flow to a new fixed point which we call ${\bf k}$
below.
Of course, for a generic value of $k$ the superpotential \wel\ is only
relevant for sufficiently large $x$, $x>x_k$.
In \KutasovIY\  $x_k$ was shown to be bounded from above,
\eqn\xkbound{  x_k<{4-\sqrt{3}\over6} (k+1).   }
This inequality is saturated in the limit $k\ra\infty$.
In the regime $x>x_k$ the $R$-charges in the fixed point  ${\bf k}$ are
fixed by the condition
\eqn\rx{   R(X) (k+1)=2.    }
Hence, the central charges can be immediately computed and no
a-maximization is required.
As before, one needs to take care of the mesons whose  $R$-charges
\eqn\mesW{
R({\cal M}_j)=2y_k+(j-1){1-y_k\over x}=2{j+k-2x\over k+1}~.
}
violate the unitarity bound.
Namely, we need to use \acorrect\ and \ccorrect\ together with
\rx\ and \mesW\ to compute the central charges.
As an example, we quote the result for $\al$:
\eqn\correctk{
\eqalign{
&\al_k(x)/N_f^2=6x(y_k-1)^3-2x(y_k-1)+3x^2\left({1-y_k\over x}-1\right)^3-x^2\left({1-y_k\over x}-1\right)+2x^2+\cr
&\cr
&{1\over 9}\sum_{j=1}^{p(x)}
\left[2- 3\left(2y_k + (j - 1){1 - y_k\over x}\right)\right]^2\left[5 - 3\left(2y_k + (j - 1){1 - y_k\over x}\right)\right]
=\cr
&{4\over (k+1)^3}\left[x^2(2+k+5k^2-12x^2)-{1\over 9}\sum_{j=1}^{p(x)}(-5+6j+k-12x)(1-3j-2k+6x)^2\right]~.}
}
Here $p(x)$ is the number of mesons which are free at $x$,
\eqn\px{
p(x)=\left\{\eqalign{&\left[{1\over 3}(6x-2k+1)\right]\quad {\rm if}\, \left[{1\over 3}(6x-2k+1)\right]\leq k\cr
&k\quad{\rm otherwise}}\right\}~,
}
where $[...]$ is the integer part of the expression in brackets if the expression is positive
and $0$ otherwise. The  $p$-th meson becomes free at
\eqn\xel{
x(p)={1\over 6}{(3p+2k-1)}~.
}
\midinsert\bigskip{\vbox{{\epsfxsize=3in
        \nobreak
    \centerline{\epsfbox{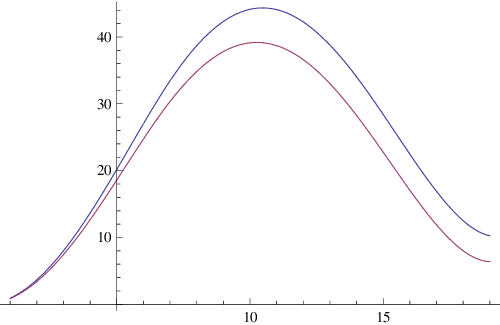}}
        \nobreak\bigskip
    {\raggedright\it \vbox{
{\bf Fig 2.}
{\it  $a(x)$ (red, bottom) and $c(x)$ (blue, top) for $sQCD$ with $W=\tr X^{21}$. 
}}}}}}
\bigskip\endinsert
\noindent
\midinsert\bigskip{\vbox{{\epsfxsize=3in
        \nobreak
    \centerline{\epsfbox{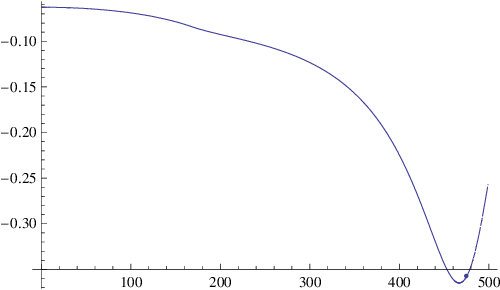}}
        \nobreak\bigskip
    {\raggedright\it \vbox{
{\bf Fig 3.}
{\it  $\alpha(x)$ for $sQCD$ with $W=\tr X^{501}$. 
}}}}}}
\bigskip\endinsert
\noindent
Taking into account the mesons corrections is important for the
positivity of $a$ and $c$.
In Fig. 2 these central charges are shown as functions of $x$.
In Fig. 3 $\alpha$ is shown as a function of $x$ for $k=500$.
Of course, these curves are only valid for $x>x_k$ and, in addition,
the region of validity is bounded from above.
However it is already clear that the bounds on $\alpha$ are not
violated.


\subsec{Strong-weak coupling duality}

\noindent
In this section, we will discuss the fixed point ${\bf k}$ obtained by perturbing adjoint
$sQCD$ by the superpotential \wel, where a dual description
is known to exist \refs{\KutasovVE\KutasovNP-\KutasovSS}, and one can ask what it
predicts for the properties of the fixed point ${\bf k}$ at strong coupling.
The duality of \refs{\KutasovVE\KutasovNP-\KutasovSS} relates adjoint $sQCD$ with
gauge group $SU(N_c)$ and superpotential \wel\  to an $\NN=1$ supersymmetric
gauge theory with gauge group $SU({\tilde N}_c)=SU(kN_f-N_c)$ and the following 
matter content: an adjoint field $Y$, $N_f$ chiral superfields $q_i$, 
$\tilde q^{\tilde i}$ in the anti-fundamental and fundamental representation of the
gauge group, respectively, and gauge singlets $(M_j)^i_{\tilde i}$, $j=1,\cdots,k-1$.
The  superpotential of this  theory is given by
\eqn\wmag{
W_{mag}=-g'_k\tr Y^{k+1}+{1\over \mu^2}g'_k\sum_{j=1}^{k}M_j\tilde q Y^{k-j}q~,
}
where $\mu$ is an auxiliary scale.
We will refer to these two  theories as electric and magnetic, respectively.
The conjecture of \refs{\KutasovVE\KutasovNP-\KutasovSS} is that they flow in the infrared
to the same fixed point. 

It will be convenient to introduce the magnetic dual of $x$,
\eqn\xequiv{
  \x \equiv { {\tilde N}_c \over N_f}=k-x~.
}
We will mainly discuss the region
\eqn\xdphys{
x, \x \in \left ({1 \over 2},k-{1 \over 2}\right)~,
}
in which both the electric and the magnetic theories are asymptotically free.

When $\x$ is close to (and above) $1/2$, most of the terms in the
superpotential \wmag\ are irrelevant. The only exception
is the term $M_k\tilde q q$, which is relevant in the infrared
fixed point of the magnetic adjoint $sQCD$ for all $\x>1/2$.
This term drives the theory to a new fixed point, at which
$M_1,\cdots, M_{k-1}$ are free but $M_k$ is interacting.
The $R$-charges and central charge $\tilde a^{m}$ at this fixed
point can be determined in a similar way to that employed in
section 2. Denoting the $R$-charge of $q,\tilde q$ by $\tilde y$, one has
$R(Y)=(1-\tilde y)/\x$ and $R(M_k)=2-2\tilde y$. Plugging
these charges into the expression for $\tilde a$, and maximizing
w.r.t. $\tilde y$, one can determine $\tilde y$ and $\tilde a^{m}$.
When $\x$ increases further, more and more of the terms in the
superpotential \wmag\ become relevant and have to be taken into
account. 
As shown in  \KutasovIY, as long as $\x<\x_k$, 
one can proceed in a way similar to that employed in the previous section.
The magnetic central charges
computed with the assumption that the last $p=(0,1,2,...)$ meson fields
$M_k,\cdots,M_{k-p+1}$ {\it are not} free are given by the expressions similar to
\acorrect\ and \ccorrect:
\eqn\acorrectm{  \al^{m,(p)}=\al^{m,(0)}+\sum_{j=1}^p \left[3 R(M_j)+R(M_j)-{2\over9}\right]+{2\over9}k   }
and
\eqn\ccorrectm{  \cl^{m,(p)}=\cl^{m,(0)}+\sum_{j=1}^p \left[3 R(M_j)+{5\over3}R(M_j)-{4\over9}\right]+{4\over9}k   }
where $\al^{m,(0)}$ and $\cl^{m,(0)}$ are given by \naiv\ and \naivc\ with the
substitution $x\ra\x$ and $y\ra\y$.
The result for $ \al^{m,(p)}$ reads
\eqn\amfna{\eqalign{
&\al^{m,(p)}/N_f^2
=6\x(\y-1)^3-2\x(\y-1)+3\x^2\left({1-\y\over \x}-1\right)^3-
\x^2\left({1-\y\over \x}-1\right)+2\x^2+\cr
&\cr
&{1\over 9}\sum_{j=1}^p
\left[2- 3\left(2\y + (j - 1){1 - \y\over \x}\right)\right]^2
\left[5 - 3\left(2\y + (j - 1){1 - \y\over \x}\right)\right]+{2\over 9}(k-2p)~.
}}
The $R$-charges can now be determined as follows.
Start with $\al^{m,(1)}$and maximize it w.r.t. $\y$. Denote the value of
$\y$ at the maximum by  $\y^{(1)}(x)$. Vary $\x$ to the point where the $R$-charge
of $M_{k-1}$ approaches $2/3$. At that point the term $M_{k-1}\tilde qYq$
in the magnetic superpotential \wmag\ becomes relevant, and one should switch to
the $\al^{m,(2)}$ description. This  can be continued to arbitrarily large
$\x$.
\midinsert\bigskip{\vbox{{\epsfxsize=3in
        \nobreak
    \centerline{\epsfbox{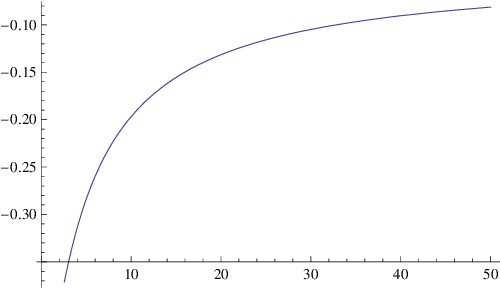}}
        \nobreak\bigskip
    {\raggedright\it \vbox{
{\bf Fig 4.}
{\it  $\alpha(\x)$ for magnetic $sQCD$ at $k=50$. 
}}}}}}
\bigskip\endinsert
\noindent
The values of $a$, $c$ and $\alpha$ can then be easily computed.
The result for the latter is shown in Fig. 4.
Apparently $\alpha$ is closest to the lower bound when the
magnetic theory is free, at $\x=1/2$.

It is interesting to analyze the theory near this point, where
only $M_k$ is not free.
In this case, maximizing \amfna\ with respect to $y$ one arrives at
\eqn\ymagn{  \y= {6\x^2-9\x-3-\sqrt{20\x^4-48\x^3+87\x^2-16\x}\over3(2\x^2-8\x-1)}.  }
One can use this result to compute $c$ and $\alpha$ at $\x=1/2$:
\eqn\alphamfree{  \alpha(\x=-{1\over2})=-{1\over2}+{3\over4 (k+2)}.   }
The lower bound is saturated in the limit $k\ra\infty$. This is because in this
limit there are infinitely many free chiral superfields which saturate the lower bound in \alphabound.

\newsec{$sQCD$ with two adjoints.}

\noindent
We turn our attention to $\NN=1$ $SYM$ theories with two adjoint chiral multiplets, which will
be denoted by $X$ and $Y$. Asymptotic freedom implies here that $x\geq 1$. Computing the $R$ charges in the $IR$  for these theories was extensively discussed in
\IntriligatorMI\  and we will closely follow this analysis. 

The simplest theory (denoted by $\hat O$ in \IntriligatorMI) is the theory with two adjoints without a superpotential. One can find the $R$-charges 
by $a$-maximization. The results are
\eqn\ORQ{\eqalign{& R(Q)=R(\tilde Q)=1+\frac{3x-2x\sqrt{26x^2-1}}{3(8x^2-1)},\cr
&R(X)=R(Y)=\half+\frac{-3+2\sqrt{26x^2-1}}{6(8x^2-1)}.}}
 There are no gauge invariant composites which violate the unitarity bound for any
value of $x$ as all the $R$ charges are larger than half. In fig. 5 we depict $\alpha(x)$  for this theory. 
\noindent
\midinsert\bigskip{\vbox{{\epsfxsize=3in
        \nobreak
    \centerline{\epsfbox{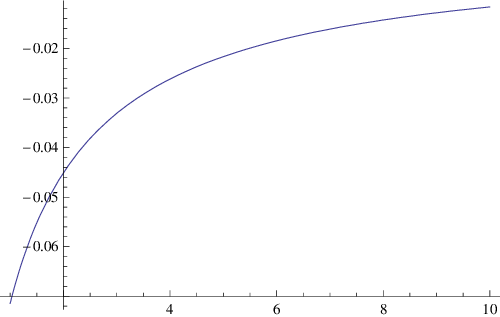}}
        \nobreak\bigskip
    {\raggedright\it \vbox{
{\bf Fig 5.}
{\it  $\alpha(x)$ for the $\hat O$ theory.
}}}}}}
\bigskip\endinsert
\noindent

 The nontrivial $IR$ fixed points of the $\NN=1$ $SYM$ theories with two adjoints and a superpotential are classified using an $ADE$-like structure
 \IntriligatorMI. The $A$ series is equivalent in the $IR$ to the theory with single adjoint discussed in the previous section,
 and we will discuss in what follows the $D$ and the $E$
 series
.

\subsec{$\hat D$.}

\noindent
We start our discussion with
the $D$ series.
 The superpotential
is given by
\eqn\Dksupot{W_{D_{k+2}}=Tr\left(X^{k+1}+XY^2\right).}
 
The first theory in this sequence is obtained by taking $k=-1$ above, and is denoted as $\hat D$.
 One can choose the following parameterization for the $R$ charges
\eqn\DRcharge{R(Q)=R(\tilde Q)=y,\qquad R(Y)={y-1\over x}+1,\qquad R(X)={2-2y\over x}.} To determine the $R$ charges there is a need for $a$-maximization.
The trial functions thus take the following form,
\eqn\atrialD{\eqalign{\tilde a^{(0)}_{\hat D}/N_f^2=&2x^2+3x^2\left({y-1\over x}\right)^3+3x^2\left({2-2y\over x}\right)^3+6x(y-1)^3 \cr
&-x^2\left({y-1\over x}\right)-x^2\left({2-2y\over x}\right)-2x(y-1).}}
\eqn\ctrialD{\eqalign{\tilde c^{(0)}_{\hat D}/N_f^2=&{4 \over 3}x^2+3x^2\left({y-1\over x}\right)^3+3x^2\left({2-2y\over x}\right)^3+6x(y-1)^3 \cr
&-{5\over3}x^2\left({y-1\over x}\right)-{5\over3}x^2\left({2-2y\over x}\right)-{10\over3}x(y-1).}}
The maximization of $\tilde a_{\hat D}$ gives 
\eqn\naiveD{y^{(0)}=1+{x(12-\sqrt{11+38x^2})\over3(2x^2-7)}.}
We have to take into account the decoupling mesons. The mesons in this theory
are $\tilde Q X^l Y^j Q$, where $j=0,1$ and $l$ is non-negative. From \naiveD\ and \DRcharge\ the only mesons
which can violate the unitarity bound have $j=0$. No baryons ever hit the unitarity bound
in the $\hat D$ theory \IntriligatorMI. The $\alpha(x)$ function for the $\hat D$ theory is depicted on fig. 6.

\midinsert\bigskip{\vbox{{\epsfxsize=3in
        \nobreak
    \centerline{\epsfbox{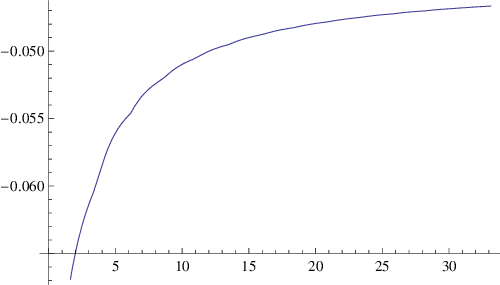}}
        \nobreak\bigskip
    {\raggedright\it \vbox{
{\bf Fig 6.}
{\it  $\alpha(x)$ for the $\hat D$ theory.
}}}}}}
\bigskip\endinsert
\noindent

\subsec{$D_{k+2}$.}

\noindent
For $k>-1$ there is no need for $a$-maximization as the charges are fixed by the marginality of the superpotential 
in the $IR$.
To actually compute $a$ we still have to take into account the decoupled free composites. 
The case of even and odd $k$ are qualitatively different. The chiral ring relations are
\eqn\Dkchiralring{\{X,Y\}=0,\qquad X^k+Y^2=0.}
For $k$ odd these imply that $Y^3=0$ and the spectrum of mesons is
truncated. For even $k$ there is no such truncation. First, we concentrate on the odd case.

  There are 
no decoupling baryons in the stability region $x<3k$ \IntriligatorMI , and thus only the mesons have to be taken into account.
The results are depicted in fig. 7 for the $\hat D_5$ theory (for all odd $k$ the results are qualitatively similar).
 The lower bound on the ratio of the central charges is saturated exactly at $x=3k$ for any odd $k$.
\midinsert\bigskip{\vbox{{\epsfxsize=3in
        \nobreak
    \centerline{\epsfbox{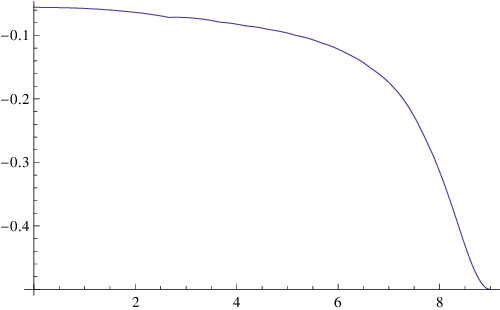}}
        \nobreak\bigskip
    {\raggedright\it \vbox{
{\bf Fig 7.}
{\it  $\alpha(x)$ for the $D_5$ theory, $k=3$.
}}}}}}
\bigskip\endinsert

However, as we approach $x=3k$ one should switch to the magnetic description of the model as it  provides us with
a more reliable picture of the physics. The magnetic dual of the $D_{k+2}$ theories was introduced in \Brodie. This theory has an $SU(3kN_f-N_c)$ gauge group, two adjoints $\tilde X$
and $\tilde Y$, $N_f$ quarks $q_i$ and $\tilde q_i$, and $3kN_f$ gauge singlets $M_{lj}$ ($l=1,\dots,3$, $k=1,2,3$). The tree
level superpotential is
\eqn\magnetDksup{W=a_0\, Tr\tilde X^{k+1}+a_1\, Tr\tilde X\tilde Y^{2}+\sum_{k,l}b_{lj}\, M_{lj}\tilde q\tilde X^{k-l}\tilde Y^{3-j}q.} 
One can compute the $a$ and $c$ charges in the magnetic theory using the techniques applied to the electric theory above \IntriligatorMI. 

Depending on the value of $x$ we should trust one of the results, magnetic or electric.
 The explicit picture is as follows.
For $x\leq 1$ the theory is free in $IR$. Beyond $x=1$ it becomes asymptotically free. For $x<x_{D_{k+2}}^{min}$ the interaction $TrX^{k+1}$ is
irrelevant and the theory flows to the $\hat D$ point. By definition at   $x=x_{D_{k+2}}^{min}$ the interaction becomes relevant and the electric
theory \Dksupot\ gives the correct physics.
 For magnetic theory it is useful to define $\tilde x=3k-x$, and the analysis above can be repeated with the magnetic dual and by exchanging $x$ with
$\tilde x$. For $\tilde x<\tilde x_{D_{k+2}}^{min}$ the magnetic theory flows to a ``magnetic'' version of the $\hat D$ point. In the conformal 
window, $x_{D_{k+2}}^{min}<x<3k-\tilde x_{D_{k+2}}^{min}$ , both descriptions should agree.  
In fig. 8 we depict  the behavior of $\alpha(x)$ near 
$\tilde x=\tilde x_{D_{k+2}}^{min}$ for $k=3$ in the magnetic theory. The conformal window begins at $\tilde x^{min}_{D_5}\sim 1.86$ ($x\sim 7.14$) and for
$x<\tilde x^{min}_{D_5}$ the electric result depicted in fig. 7 can be trusted.
\midinsert\bigskip{\vbox{{\epsfxsize=3in
        \nobreak
    \centerline{\epsfbox{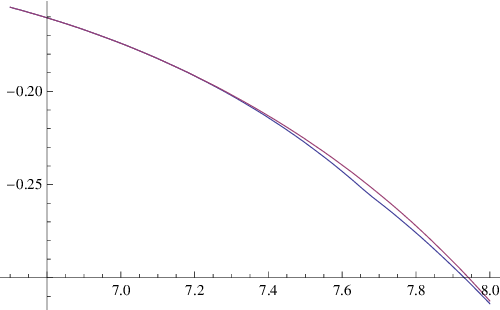}}
        \nobreak\bigskip
    {\raggedright\it \vbox{
{\bf Fig 8.}
{\it  $\alpha(x)$ for the magnetic (top, red) and electric (bottom, blue) $D_5$ theories.
}}}}}}
\bigskip\endinsert
\noindent

One can repeat the analysis above for the case of even $k$. In figures 9 and 10 we depict the results for $k=2$ and $k=4$.
\midinsert\bigskip{\vbox{{\epsfxsize=3in
        \nobreak
    \centerline{\epsfbox{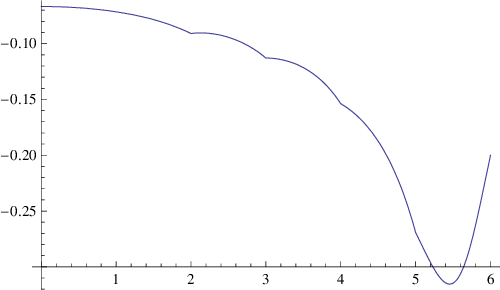}}
        \nobreak\bigskip
    {\raggedright\it \vbox{
{\bf Fig 9.}
{\it  $\alpha(x)$ for the  $D_2$ theory.
}}}}}}
\bigskip\endinsert
\noindent
\midinsert\bigskip{\vbox{{\epsfxsize=3in
        \nobreak
    \centerline{\epsfbox{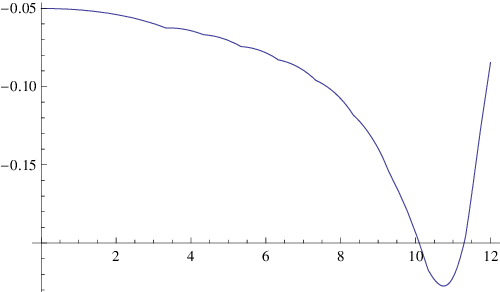}}
        \nobreak\bigskip
    {\raggedright\it \vbox{
{\bf Fig 10.}
{\it  $\alpha(x)$ for the  $D_4$ theory.
}}}}}}
\bigskip\endinsert
\noindent

\subsec{$\hat E$.}

\noindent
Next we turn our attention to the $E$ series.
The $E$ series is defined by the  superpotential $W_{\hat E}=Tr\, Y^3$ and the following deformations,
\eqn\Esupot{W_{E_6}=Tr\left(X^4+Y^3\right),\qquad W_{E_7}=Tr\left(Y\,X^3+Y^3\right),\qquad W_{E_8}=Tr\left(X^5+Y^3\right).}
We start by considering the basic case $W_{\hat E}$. The $R$-charges are parameterized as
\eqn\ER{R(Y)={2\over 3},\qquad R(Q)=R(\tilde Q)=y,\qquad R(X)={1+x-y\over x}-{2 \over 3}.}
As we have a free parameter $y$ we have to use $a$-maximization. Here we do not have any gauge invariant operators
violating the unitarity bound and thus the calculation is straightforward. In fig. 11 we depict the relevant diagram.
\midinsert\bigskip{\vbox{{\epsfxsize=3in
        \nobreak
    \centerline{\epsfbox{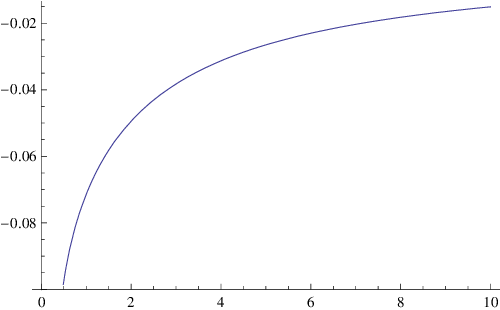}}
        \nobreak\bigskip
    {\raggedright\it \vbox{
{\bf Fig 11.}
{\it  $\alpha(x)$ for the $\hat E$ theory. 
}}}}}}
\bigskip\endinsert
\noindent

\subsec{$E_6$.}

\noindent
There are three interesting deformations of the $\hat E$ theory. We start with the $E_6$ point. Here the $R$-charges
are fixed by demanding marginality of the superpotential,
\eqn\ERsix{R(Y)={2\over 3},\qquad R(Q)=R(\tilde Q)=1-{x \over 6},\qquad R(X)={1\over 2}.}
However we have here decoupling mesons (and baryons). A meson with $k$ $Y$ fields and $l$ $X$ fields decouples at
\eqn\Esixmeson{x=4+{3\over 2}l+2k.} One has to account for the different mesons keeping in mind the chiral ring 
identifications $Y^2=0$ and $X^3=0$. In fig. 12 we depict the results. 
\midinsert\bigskip{\vbox{{\epsfxsize=3in
        \nobreak
    \centerline{\epsfbox{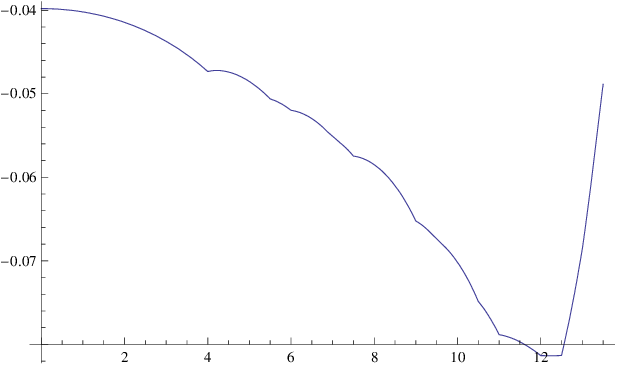}}
        \nobreak\bigskip
    {\raggedright\it \vbox{
{\bf Fig 12.}
{\it  $\alpha(x)$ for the $E_6$ theory. The theory is conjectured to possess a quantum instability at roughly $x=13.8$. 
}}}}}}
\bigskip\endinsert

\subsec{$E_7$.}

\noindent
Let us discuss the  $E_7$ point. Here the $R$-charges
are fixed as,
\eqn\ERsix{R(Y)={2\over 3},\qquad R(Q)=R(\tilde Q)=1-{x \over 9},\qquad R(X)={4\over 9}.}
we have here decoupling mesons (and baryons). A meson with $k$ $Y$ fields and $l$ $X$ fields decouples at
\eqn\Esevenmeson{x=6+2l+3k.} One has to account for the different mesons keeping in mind the chiral ring 
identifications $\{Y,X^2\}=0$ and $X^3+3Y^2=0$. In fig. 13 we depict the results. 
\noindent
\midinsert\bigskip{\vbox{{\epsfxsize=3in
        \nobreak
    \centerline{\epsfbox{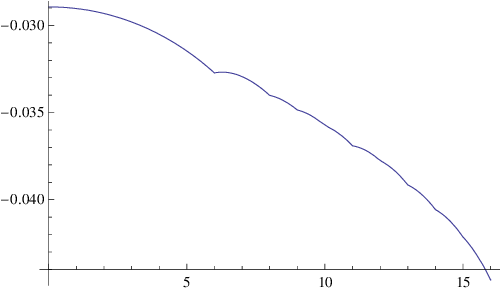}}
        \nobreak\bigskip
    {\raggedright\it \vbox{
{\bf Fig 13.}
{\it  $\alpha(x)$ for the $E_7$ theory. 
}}}}}}
\bigskip\endinsert
\noindent

\subsec{$E_8$.}

\noindent
Finally we discuss the $E_8$ case.
The superpotential has the form $Tr\left(X^5+Y^3\right)$. This implies the chiral ring relations
$X^4=0$ and $Y^2=0$. The $R$ charges are as follows 
\eqn\REeight{R(Q)=R(\tilde Q)=1-{x\over 15},\qquad R(X)={2\over 5},\qquad R(Y)={2\over 3}.}
There is no $a$-maximization but we have to take into account decoupling mesons and baryons. The lowest lying mesons
are summarized below,
\eqn\EeightMes{\eqalign{&\left(\tilde QQ,\, 10\right),\quad \left(\tilde QXQ,\, 13\right),\quad \left(\tilde QYQ,\, 15\right),\quad
\left(\tilde QX^2Q,\, 16\right),\quad \left(\tilde QXYQ,\,\tilde QYXQ,\, 18\right),\cr
&\left(\tilde QX^3Q,\, 19\right),\quad \left(\tilde QYX^2Q,\,\tilde QXYXQ,\,\tilde QX^2YQ,\, 21\right),\quad
\left(\tilde QYXYQ,\, 23\right),}} where we also included the value of $x$ for which the mesons decouple.
The baryons are built from the dressed quarks
\eqn\EeightQ{\hat Q^{(n)}_{\beta,\alpha,\gamma}=X^{\beta}\left[\prod_{i=1}^nYX^{\alpha_i}\right]Y^\gamma Q,\qquad \alpha_i=1,2,3\quad \gamma=0,1 \quad
\beta=0,1,2,3.}
Thus the baryons are products of $N_c$ of these dressed quarks. For sufficiently large values of $x$ the baryons will decouple
and the proof is the same as for the $E_6$ case \IntriligatorMI. We can find lower bounds on the values of $x$ for the baryon decoupling, $x_\star$.
The $R$ charge of any baryon is greater than
\eqn\EeightBound{N_c(1-{x\over 15})+{2\over 5}(N_c-N_f),} just by taking $N_f$ un-dressed quarks
and the rest being dressed with $X$. Thus,
we get a very rough lower bound of $x_\star > 20$.   
\noindent
\midinsert\bigskip{\vbox{{\epsfxsize=3in
        \nobreak
    \centerline{\epsfbox{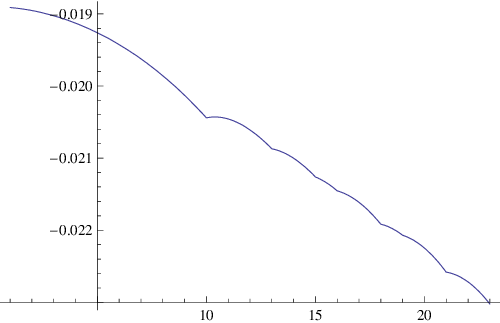}}
        \nobreak\bigskip
    {\raggedright\it \vbox{
{\bf Fig 14.}
{\it  $\alpha(x)$ for the $E_8$ theory. 
}}}}}}
\bigskip\endinsert
\noindent

\newsec{Conclusions.}

In this note we have computed the value of $\alpha=a/c-1$ for a wide range 
of interacting $\NN=1$ superconformal theories. We have verified that
the bound \alphabound\ suggested by Hofman and Maldacena in \HofmanAR\ is satisfied in these theories.
It is interesting to note that in all the
interacting cases considered  $\alpha$ is actually negative
(and larger than $-1/2$).
It is also worth noting that for interacting theories with gravity duals, which
presumably are strongly coupled,  $\alpha=0$.
On the other hand, in our examples strong coupling regime, which would happen in 
the middle of the conformal window, was not associated with small values
of $\alpha$.
As far as outlook for the future is concerned, it would perhaps be 
interesting to consider more examples of $CFT$s in order to understand better the
significance of $\alpha$.
It would also be great to have a proof of  \alphabound\ and of 
its non-supersymmetric cousin.

\bigskip
\bigskip

\noindent {\bf Acknowledgements:}
We thank O.~Aharony, D.~Kutasov, J.~Maldacena and A.~Shapere for very useful
discussions and comments on the manuscript. This research is supported in part by 
the National Science Foundation Grant No.
PHY-0653342.

\listrefs
\end